\documentclass[fleqn,10pt]{wlscirep}
\title{Quantum Monte Carlo simulation of a particular class of non-stoquastic Hamiltonians in quantum annealing}

\author[1,*]{Masayuki Ohzeki}
\affil[1]{Graduate School of Information Sciences, Tohoku University, Sendai 980-8579, Japan}

\affil[*]{mohzeki@smapip.is.tohoku.ac.jp}

\affil[+]{The author has carried out all the work alone}

\begin{abstract}
Quantum annealing is a generic solver of the optimization problem that uses fictitious quantum fluctuation.
Its simulation in classical computing is often performed using the quantum Monte Carlo simulation via the Suzuki--Trotter decomposition.
However, the negative sign problem sometimes emerges in the simulation of quantum annealing with an elaborate driver Hamiltonian, since it belongs to a class of non-stoquastic Hamiltonians.
In the present study, we propose an alternative way to avoid the negative sign problem involved in a particular class of the non-stoquastic Hamiltonians.
To check the validity of the method, we demonstrate our method by applying it to a simple problem that includes the anti-ferromagnetic XX interaction, which is a typical instance of the non-stoquastic Hamiltonians.
\end{abstract}
\begin{document}

\flushbottom
\maketitle

\thispagestyle{empty}

\section*{Introduction}
Quantum annealing (QA) is a generic algorithm aimed at solving optimization problems by exploiting the quantum tunneling effect.
The scheme was originally proposed as a numerical computational algorithm \cite{Kadowaki1998} inspired by simulated annealing (SA) \cite{Kirkpatrick1983}, and the exchange Monte Carlo simulation \cite{Hukushima1996}.
In QA, the quantum tunneling effect efficiently finds the ground state even in the many-valley structure of the energy landscape therein \cite{Ray1989,Apolloni1989,Arnab2008}.
The challenge for searching the ground state for the NP-hard problem has been also attempted in the literature \cite{Das2008}.
The extremely slow driving of the quantum fluctuations keeps the system in the ground state of the instantaneous Hamiltonian due to adiabacity.
The concept of adiabacity is essentially utilized in QA as reported in the literature \cite{Farhi2001}. 
Indeed, the theoretical aspects of QA are well known and its basic concept owes its origin to the quantum adiabatic theorem \cite{Suzuki2005,Morita2008,Ohzeki2011c}.
The successful implementation of QA demands highly controllable engineering on qubits, but its experimental realization has been accomplished recently by the use of current technology \cite{Dwave2010a,Dwave2010b,Dwave2010c,Dwave2014}, and is rapidly gathering considerable attention.
Before appearance of the D-Wave machines, the experimental realization of QA could be found in determining the ground state for some molecules with complex structures \cite{Finnila1994}.
In several protocols based on QA, they do not necessarily adhere to the adiabatic quantum computation or keep the system in the ground state but instead also employ the non-adiabatic counterpart \cite{Ohzeki2010a,Ohzeki2011,Ohzeki2011proc,Somma2012}.
As per a recent study on the performance of QA by using the D-wave machine, the experimental behavior can be indeed understood by thermalization and non-adiabatic behavior of the quantum system \cite{Amin2015}.

In QA, we formulate the platform to solve the optimization problem using the time-dependent Hamiltonian.
The initial Hamiltonian is governed by the ``driver" Hamiltonian considering full quantum fluctuations.
The often-used example is the transverse field.
The first stage of QA is initialized initialization of the ground state of the driver Hamiltonian.
The quantum effect will be gradually turned off, only with the classical Hamiltonian remaining at the end of the algorithm.
The Hamiltonian takes the form of the classical Ising model at the final time.
Numerous reports state that the performance of QA is better than that of SA.
This may be because the quantum tunneling effect penetrates the valley of the potential energy.
The QA system with the transverse field is in a class of the stoquastic Hamiltonian, which can be straightforwardly simulated by use of the quantum Monte Carlo simulation in a classical computer \cite{Santoro2006}.
The other type of quantum driver may be used for improving the performance of quantum annealing, a successful instance of which is the implementation of the anti-ferromagnetic XX interaction \cite{Seki2012,Seki2015}.
However, this is a typical example of the non-stoquastic Hamiltonian, which cannot be straightforwardly simulated in a classical computer.
The definition of the non-stoquastic Hamiltonian is that the instantaneous Hamiltonian has non-positive (negative semi-definite) elements in the non-diagonal elements \cite{Bravyi2008}.
Thus, the naive application of the quantum Monte Carlo simulation via the Suzuki--Trotter decomposition \cite{Suzuki1976} or various kinds of the classical-quantum mapping \cite{Castelnovo2005,Somma2007,Nishimori2015} suffers from the negative sign problem.
The negative sign problem degrades the precision for estimation of the thermodynamic quantities in the numerical stochastic computation of the quantum Monte Carlo simulation.
To remove the negative sign problem in the numerical computation of the non-stoquastic Hamiltonian, one seeks a proper basis to represent the Hamiltonian without the negative value in the non-diagonal elements \cite{Nakamura1998,Harada2014}.
However, this task is highly nontrivial and there are no generic solutions.
The negative sign problem belongs to the class of the NP-hard problem \cite{Matthias2005}.

In the present study, we show a method to simulate a particular class of the non-stoquastic Hamiltonians including the anti-ferromagnetic XX interaction in the quantum Monte Carlo simulation.
The numerical scheme will be demonstrated below.
The present study paves the way to simulate quantum annealing with an elaborate driver Hamiltonian.
The future development of hardware devices to perform quantum annealing aims at implementing the non-stoquastic Hamiltonian such as the anti-ferromagnetic XX interaction beyond the classically simulatable world.
Our contribution in the present study is to establish a test bed to simulate a particular class of non-stoquastic Hamiltonians in order to verify the performance of future hardware.
Moreover, our study will enable invention of many types of algorithms inspired by quantum annealing in the classical computer but with the non-stoquastic Hamiltonian.

In standard QA, we employ a system with the transverse field as
\begin{equation}
H_1({\boldsymbol \sigma}) = H_0({\boldsymbol \sigma}) - \Gamma\sum_{i=1}^N \sigma_{i}^x,
\end{equation}
where $\Gamma$ represents the strength of the transverse field, and $\sigma_{i}^x$ is the $x$ component of the Pauli matrix.
The classical Ising model to be solved is $H_0({\boldsymbol \sigma})$, where ${\boldsymbol \sigma} = (\sigma^z_1,\sigma^z_2,\cdots,\sigma_N^z)$.
 
The standard QA procedure decreases the strength of the quantum driver Hamiltonian to find the ground state of the cost function to be optimized.
The adiabatic theorem ensures that we obtain the ground state after sufficient slow sweep of the quantum driver Hamiltonian.
The computational time of QA can be evaluated by the energy gap between the ground state and the first excited state during the decrease in the transverse fields \cite{Suzuki2005,Morita2008}.
As has been extensively studied, the quantum phase transition hampers the efficient computation of QA.
In particular, the first-order phase transition characterized by the exponential closure of the energy gap depending on the number of spins signals the long-time computation for optimization via QA.
Seki and Nishimori successfully avoided the first order phase transition by utilizing additional fluctuations, anti-ferromagnetic XX interactions.
\begin{equation}
H_2({\boldsymbol \sigma}) = H_0({\boldsymbol \sigma}) - \Gamma\sum_{i=1}^N \sigma_{i}^x + \frac{\gamma}{N} \left(\sum_{i=1}^N \sigma_{i}^x\right)^2,\label{H2}
\end{equation}
However, their quantum fluctuation yields the negative sign problem in a naive application of the Suzuki--Trotter decomposition in the quantum Monte Carlo method \cite{Suzuki1976}.
Seki and Nishimori avoid the negative sign problem without rotating the basis to represent the Hamiltonian.
They utilize the mean-field analysis to compute the free energy and the thermodynamic quantities such as the magnetization, energy, etc.
A recent study on the performance in the optimization problem by use of the anti-ferromagnetic XX interaction is carried out by the numerical diagonalization for a limited size of the system\cite{Hormozi2016}.

We then elucidate the essential part of the mean-field study to avoid the negative sign problem and apply its scheme to simulate the non-stoquastic Hamiltonian with a more generic form. 
The scheme involves the following form of the non-stoquastic Hamiltonian 
\begin{equation}
H({\boldsymbol \sigma}) = H_0({\boldsymbol \sigma}) - N f\left( \frac{1}{N}\sum_{i=1}^N \sigma_{i}^x \right),\label{NonStoQA}.
\end{equation}
The classical Hamiltonian here is not restricted to the all-to-all connections as in the previous studies carried out by Seki and Nishimori \cite{Seki2012,Seki2015}.
The proposed method in the present study is applicable to all kinds of the classical Hamiltonian.
For instance, the spin-glass Hamiltonian with all-to-all connections as
\begin{equation}
H_0({\boldsymbol \sigma}) = - \sum_{i\neq j}J_{ij} \sigma_i^z \sigma_j^z - \sum_{i=1}^N h_i \sigma_i^z,
\end{equation}
and the finite-dimensional spin-glass Hamiltonian with limited connections as
\begin{equation}
H_0({\boldsymbol \sigma}) = - \sum_{\langle i j \rangle}J_{ij} \sigma_i^z \sigma_j^z - \sum_{i=1}^N h_i \sigma_i^z,
\end{equation}
where, $\langle ij \rangle$ stands for the summation over the subset of the interactions, which are located at each bond on the specific lattice.
The dimensionality is not limited because our method is based on the quantum Monte Carlo simulation rather than the mean-field analyses as in the previous studies \cite{Seki2012,Seki2015}.
In addition, the various forms of the Hamiltonian representing the optimization problem with many-body interactions can be within the scope of application of our method.
On the other hand, the quantum fluctuation term is limited to the specific form, in which its argument is given by the summation over all-components of the $x$-component of the Pauli operators.
This is a kind of extension of the quantum fluctuation used in QA.
The case of the standard QA is described by $f(m_x) =  \Gamma m_x$.
The case of the QA with the anti-ferromagnetic XX interaction and transverse field is included in this form through $f(m_x) = \Gamma m_x - \gamma m_x^2/2$.

In our scheme, we utilize the quantum Monte Carlo simulation while avoiding the negative sign problem for a particular class of the non-stoquastic Hamiltonian through the simple calculations by replacement of the quantum fluctuation with the adaptive-changing transverse field.
As discussed above, there is no generic solution for the negative sign problem.
To avoid the obstacle in the numerical computation, one needs to find an adequate basis in which the negative sign problem disappears.
Contrasting with the ordinary approach overcoming the difficulty involved in the negative sign problem, our method avoids the obstacle via the simple transformation and calculations rather than finding the proper basis to represent the Hamiltonian. 
We propose two approaches to simulate a particular class of the non-stoquastic Hamiltonian.
The first one is called the adaptive quantum Monte Carlo simulation and the other is the data-analysis approach.
These methods are detailed below.

The following section demonstrates the validity of our proposed method for a simple model: the (ferromagnetic) infinite-range model with the transverse field and anti-ferromagnetic XX interaction defined as
\begin{equation}
H_0({\boldsymbol \sigma}) = - h \sum_{i=1}^N \sigma_i^z - N \left( \sum_{i=1}^N \sigma_i^z \right)^2, \label{SM}
\end{equation}
and $f(m_x) = \Gamma m_x - \gamma m_x^2/2 $.
The infinite model has the second-order phase transition at the critical point $\Gamma_c = 1$ without any (longitudinal) magnetic field and anti-ferromagnetic XX interactions.
The model we dealt with here is solvable by mean-field analysis and thus adequate for validation of our method.
We emphasize, however, that the application range of our method is beyond that of the mean-field analysis.
When we perform the naive application of the quantum Monte Carlo simulation through the standard approach via the Suzuki--Trotter decomposition, the numerical simulations suffer from the negative sign problem to obtain precise estimations of the physical quantities for the case away from $f(m_x) = \Gamma m_x$.
However, we do not encounter the negative sign problem by the use of the replacement of the non-stoquastic terms in Eq. (\ref{NonStoQA}) via the saddle-point method and integral representation of the delta function as detailed below.

\section*{Results}
\begin{center}
\begin{figure}[t]
\begin{center}
\includegraphics[width=1.05\columnwidth]{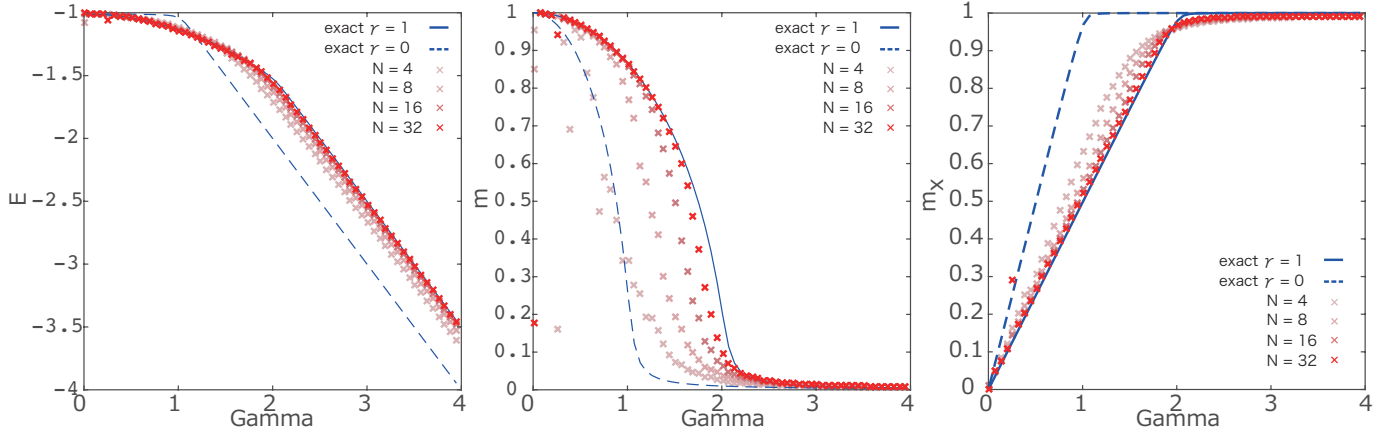}
\end{center}
\caption{Results of the adaptive quantum Monte Carlo simulation.
The dashed curves denote the exact solution for the case without antiferromagnetic interactions and solid ones represent those with antiferromagnetic interactions.
The gradating red dots represent the results of the adaptive quantum Monte Carlo simulation (denser red point stands for larger size of system). 
}
\label{result1}
\end{figure}
\end{center}

In this section, we propose our first method: the adaptive quantum Monte Carlo simulation.
We set the spin systems with $N=4,8,16$ and $32$.
In order to perform the quantum Monte Carlo simulation, we consider the replicated system at extremely low temperatures via the Suzuki--Trotter decomposition.
We set the Trotter number as $\tau=128$ and the inverse temperature (temperature) as $\beta = 50~(T=0.02)$.
To check the precision of the computation, we estimate various thermodynamic quantities: the internal energy, (longitudinal) magnetization, and the transverse magnetization including the order parameters of the system.
In Fig. \ref{result1}, we show estimations for three thermodynamic quantities by computing $80000$ MCS time average after $20000$ MCS equilibration.
For comparison, we show the exact solution carried out using the mean-field analysis, because this model is tractable.
The case with $\gamma = 1$ is for the anti-ferromagnetic XX interaction, while that with $\gamma=0$ is for absence of the anti-ferromagnetic XX interaction.
The results obtained by the adaptive quantum Monte Carlo simulation are close to the exact result with $\gamma=1$ rather than that with $\gamma=0$.
This fact is evidence that the adaptive quantum Monte Carlo simulation can estimate correctly the thermodynamic quantities even in the non-stoquastic Hamiltonian.
Aside from the finite-size effect, the obtained results asymptotically coincide with the exact solutions.
We remark that the convergence of the estimations is relatively slow compared to the ordinary quantum Monte Carlo simulation because we have to estimate the expectation of the thermodynamic quantities as the effective transverse field $f'(m_x)$ depending on the tentative value of the transverse magnetization.
Although substantial slowing down of the equilibration is not observed in the simple model, the computational time to correctly estimate the thermodynamic quantities might become longer depending on the model that is related to its complexity.

\begin{center}
\begin{figure}[t]
\begin{center}
\includegraphics[width=1.05\columnwidth]{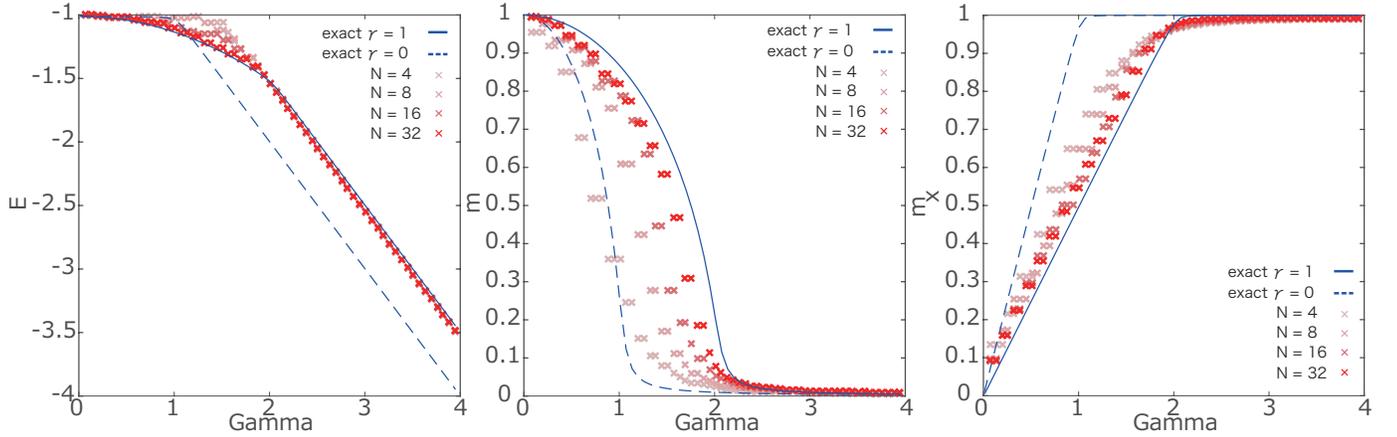}
\end{center}
\caption{Results of data analysis.
The same symbols are used as those in Fig. \ref{result1}.}
\label{result2}
\end{figure}
\end{center}

Another method to estimate the thermodynamic quantities of the non-stoquastic Hamiltonian is the data analysis approach.
Similar to the previous check, we test our method in the simple infinite-range model defined by Eq. (\ref{SM}).
We again set the spin systems with $N=4,8,16$ and $32$, the Trotter number as $\tau=128$, and the inverse temperature (temperature) as $\beta = 50~(T=0.02)$.
We here perform the ``standard" quantum Monte Carlo simulation for the model only with the transverse field while changing the value of the transverse field from $\Gamma=0$ to $\Gamma=4$ by $\Delta \Gamma = 0.05$.
For validation of our method, we estimate the internal energy, (longitudinal) magnetization, and the transverse magnetization as in Fig. \ref{result2}.
The computational time is short because we perform the ``standard" quantum Monte Carlo simulation.
We take $80000$ MCS time average after $10000$ MCS equilibration.
Similar to the results of the adaptive quantum Monte Carlo simulation, we confirm the validity of our method within the finite-size effects of the simulation.
The obtained results become closer to the exact solution with $\gamma=1$, which is the non-stoquastic Hamiltonian, rather than with $\gamma=0$.
In the results obtained using the data-analysis approach, we find several adjacent points in the different value of $\gamma$.
This means that the same results are obtained in several cases after estimation of the thermodynamic quantities for the non-stoquastic Hamiltonian.
We can remove lack of numerical precision by increasing the step size of the transverse magnetic field in the ``standard" quantum Monte Carlo simulation.
We confirm that both the proposed methods based on the quantum Monte Carlo simulation can estimate the thermodynamic quantities.

\section*{Discussion}
We propose two methods to simulate the non-stoquastic Hamiltonian on the basis of the quantum Monte Carlo simulation, which is often affected by the negative sign problem due to the computational basis.
Although we usually consider the rotation of the basis to remove the negative sign problem, we avoid the negative sign problem by introducing the auxiliary variable and its conjugate one $m_x$ and $\tilde{m}_x$.
The model which we deal with is non-stoquastic but the negative sign problem is not insurmountable.
Thus, we avoid the negative sign problem though the model is in the definition of the non-stoquastic Hamiltonian.

We list the possible questions and answers on our proposed method below:
\begin{itemize}
\item Our stand point.
In the present study, we simply check the validity of our proposed methods by comparing the numerical results and the exact solution via the mean-field analysis.
Since, the proposed method considers a toy model, we do not obtain any nontrivial result. 
However, this means readily that our work is trivial.
This is the first step to establish a systematic approach for the simulation of the non-stoquastic Hamiltonian.
In the numerical simulation of the non-stoquastic Hamiltonian with a large number of components, efficient methods to compute the thermodynamic quantities are the exact diagonalization of the Hamiltonian and the renormalization group analysis.
Both the methods often suffer from size limitation.
We propose an alternative choice to approach the non-stoquastic Hamiltonian.

\item Limitation of our method.

Our method is not a generic solution to the negative sign problem.
The applicable scope of our method is limited to the case with the quantum fluctuation determined by the collection of the $x$-component of the Pauli operators.
The form of the function $f(m)$ is very flexible.
The condition of the function is that $P(m_x) \to 0$ as $m_x \to \pm \infty$.
Furthermore, when the spin operators characterizing the quantum fluctuation can be reduced to a few quantities, our proposed method can be applied.
For instance, our method can be generalized to the case with $\sum_{p=1}^P \gamma_p f_p\left( \sum_{i \in \Lambda_p} \sigma_i^x \right)$, where $\Lambda_p$ is a subset of the indices (locations) of the spin operators.
This means that the inhomogeneous XX interactions can be dealt with.
The combination of the inhomogeneous XX interaction is directly related to the computational cost.
We would expect that the boundary between the classical and the quantum computational capacity lies in the inhomogeneity of the XX interactions.
This problem is presently not solvable.

\item Saddle point 
We employ the saddle-point method for determining the value of $\tilde{m}_x$ as detailed below.
Thus, in order to enhance the numerical precision, we take the system size to be relatively large.
However, we may simulate the value of $m_x$ using the Langevin stochastic process.
Then, we need to choose the direction of $m_x$ for stabilization of its behavior depending on the form of $f(m_x)$.

\item Capability of acceleration.

Most of the ordinary simulations by the Langevin stochastic process and the Markov-chain Monte Carlo simulation require detailed balance condition.
Various studies have ensured that the violation of the detailed balance condition exhibits acceleration of the convergence to the pre-determined steady state \cite{Ichiki2013,Ichiki2015,Ohzeki2015,Ohzeki2015proc}.
We may also utilize the exchange Monte Carlo simulation to accelerate the convergence to the steady state \cite{Hukushima1996}.
In the present study, we employ this method to obtain our results.

\item Avoiding the first-order phase transition.

Seki and Nishimori demonstrated the possibility to remove the first-order phase transition by using the anti-ferromagnetic XX interaction.
The effect of a kind of the quantum fluctuation can be recast as the modified transverse field
depending on the tentative value of the transverse magnetization.
This different approach provides a deep understanding of the change of the phase transition by using the elaborate quantum fluctuations.
As shown above, the resulting thermodynamic quantity can be characterized by the cross points between the transverse magnetization under the effective transverse field and the inverse function determined by the form of the quantum fluctuation $f(m_x)$.
The first-order phase transition is involved in the discontinuous jump of the thermodynamic quantity.
The discontinuous jump emerges from co-existing phases at the same parameter. 
In other words, the thermodynamic quantity is described by a multivalued function.
The thermodynamically stable point is selected by comparing the values of the free energy at the multiple solutions at the same parameter.
When we employ a nontrivial function as $f(m_x)$ to avoid multiple cross points, we convert the first-order phase transition to the second-order (continuous) phase transition.
The relationship between the computational complexity of QA with non-stoquastic Hamiltonian and the limitation of our approach would be a very exciting realm and will be reported elsewhere as a future study.
\end{itemize}

A recent study on the non-stoquastic Hamiltonian is carried out by numerical diagonalization for a kind of the spin glass model with long-range interaction with the anti-ferromagnetic XX interactions.
However, if one implements our method, one can investigate the property of the model.
In this sense, our methods open an alternate way to approach nontrivial aspects of the non-stoquastic Hamiltonian by performing the quantum Monte Carlo simulation for large-component systems as compared to the sizes considered in the numerical diagonalization.
We will report such nontrivial results by use of our method in the near future.

\section*{Methods}
In the standard approach of the quantum Monte Carlo simulation, we construct the replicated partition function via the Suzuki--Trotter decomposition to reduce the transverse field to the interactions between the replicated systems.
For a class of the non-stoquastic Hamiltonian written as in Eq. (\ref{NonStoQA}), we replace the quantum fluctuation with the adaptively-changing transverse field via the integral representation of the delta function and saddle-point method.
Thus, the method we detail below can be more precise when the number of degrees of freedom is large.

The partition function of the present system is written as follows:
\begin{equation}
Z = {\rm Tr}_{\boldsymbol \sigma}\left\{ \exp\left( - \beta H_0({\boldsymbol \sigma}) +N \beta f \left(\sum_{i=1}^N \sigma_{i}^x\right)\right) \right\}.
\end{equation}
We employ the Suzuki--Trotter decomposition to divide the exponentiated Hamiltonian into the diagonal and non-diagonal parts as
\begin{equation}
Z \approx \lim_{\tau \to \infty} {\rm Tr}_{\{{\boldsymbol \sigma}_t\}}\left\{ \prod_{t=1}^{\tau} e^{ - \frac{\beta}{\tau} H_0({\boldsymbol \sigma}_t)}e^{N \frac{\beta}{\tau} f \left(\sum_{i=1}^N \sigma_{it}^x\right)} \right\}.
\end{equation}
Here, we utilize an identity through the delta function as
\begin{equation}
1 = \int dm_{xt} \delta\left( m_{xt} - \frac{1}{N} \sum_{i=1}^N \sigma_{it}^x\right)
\end{equation}
The partition function is written as 
\begin{eqnarray}
Z &\approx& \lim_{\tau \to \infty} {\rm Tr}_{\{{\boldsymbol \sigma}_t\}}\left\{ \int  \prod_{t=1}^{\tau}dm_{xt} e^{ - \frac{\beta}{\tau} H_0({\boldsymbol \sigma}_t)}e^{N\frac{\beta}{\tau} f \left(m_{xt}\right)}\delta\left( m_{xt} - \frac{1}{N} \sum_{i=1}^N \sigma_{it}^x\right)\right\}.
\end{eqnarray}
This is a kind of micro-canonical form of the present system.
We further employ the Fourier integral form of the delta function as
\begin{eqnarray}
Z &\approx& \lim_{\tau \to \infty} {\rm Tr}_{\{{\boldsymbol \sigma}_t\}}\left\{ \int \prod_{t=1}^{\tau}dm_{xt}d\tilde{m}_{xt}  e^{ - \frac{\beta}{\tau} H_0({\boldsymbol \sigma}_t)}e^{N\frac{\beta}{\tau} f \left(m_{xt}\right)} e^{- \frac{\beta}{\tau}\tilde{m}_{xt}\left(Nm_{xt} - \sum_{i=1}^N \sigma_{it}^x\right)}\right\}.
\end{eqnarray}
The resulting partition function is the same as the Ising model with the transverse field.
In the thermodynamic limit, we may take the saddle point in the integral.
The saddle point is evaluated by $\tilde{m}_{xt} = f'(m_{xt})$, where $f'(m)$ is the derivative of the function $f(m)$.
In other words, the transverse field is determined by the auxiliary variable $m_{xt}$, which corresponds to the transverse magnetization.
The transverse field can be rewritten as the interaction between the different Trotter slices $t$ and $t+1$, according to the prescription of the quantum Monte Carlo simulation as
\begin{eqnarray}
Z &\approx& \lim_{\tau \to \infty} {\rm Tr}_{\{{\boldsymbol \sigma}_t\}}\left\{ \int  \prod_{t=1}^{\tau} dm_{xt} e^{ - \frac{\beta}{\tau} H_0({\boldsymbol \sigma}_t) + B_t \sum_{i=1}^N \sigma_{i,t}\sigma_{i,t+1}} e^{- N\frac{\beta}{\tau}\left(m_{xt} f'(m_{xt}) - f(m_{xt})\right)}\right\}.
\end{eqnarray}
where $B_t = - \log \tanh(\beta f'(m_{xt}))/2$.
For simplicity, we employ the static approximation as $m_{xt} = m_x$ and $\tilde{m}_{xt} = \tilde{m}_x$ below.
We define the joint probability distribution for $(\{{\boldsymbol \sigma}_t\},m_x)$ as
\begin{eqnarray}
P({\boldsymbol \sigma},m_x) &=& \frac{1}{Z}\prod_{t=1}^{\tau} e^{ - \frac{\beta}{\tau} H_0({\boldsymbol \sigma}_t) + B \sum_{i=1}^N \sigma_{i,t}\sigma_{i,t+1}} e^{- N\beta\left(m_{x} f'(m_{x}) - f(m_{x})\right)}.
\end{eqnarray}
Simultaneously, we define the conditional probability as
\begin{eqnarray}
P({\boldsymbol \sigma}|m_x) &=& \frac{1}{Z(m_x)}\prod_{t=1}^{\tau} e^{ - \frac{\beta}{\tau} H_0({\boldsymbol \sigma}_t) + B \sum_{i=1}^N \sigma_{i,t}\sigma_{i,t+1}}.\label{QMC}
\end{eqnarray}
where $Z(m_x)$ is the normalization constant and the marginal distribution as
\begin{eqnarray}
P(m_x) &=& \frac{Z(m_x)}{Z} \times e^{- N\beta\left(m_{x} f'(m_{x}) - f(m_{x})\right)}.
\end{eqnarray}

In the standard procedure of the Markov-chain Monte Carlo simulation, we generate a series of realizations of the Ising variable following the probability distribution conditioned on some fixed parameters such as the temperature and transverse field.
However, in our case, the effective transverse field can change the tentative value of the auxiliary variable as $\tilde{m}_x = f'(m_x)$.
The argument of the function $f'(m_x)$ also fluctuates following the marginal distribution. 
We may then simulate the stochastic process governed by the Langevin equation to generate the auxiliary variable $m_{x}$ following the marginal distribution as
\begin{equation}
dm_x = Nf''(m_x)\left(m_x  - \left\langle \sigma_i^x \right\rangle \right)dt + \sqrt{\frac{2}{\beta}}dW,
\end{equation}
where $dW$ is the Weiner process.
Here, we define
\begin{equation}
 \left\langle \sigma_i^x \right\rangle = \left\langle\frac{1}{\tau}\sum_{t=1}^{\tau}\tanh\left(\frac{\beta}{\tau}\tilde{m}_x\right)^{\sigma_{i,t}\sigma_{i,t+1}}\right\rangle, \label{tra}
\end{equation}
where the bracket denotes the expectation with respect to the weight of the conditional probability $P({\boldsymbol \sigma}|m_x)$.
In the case of the number of the spin variables $N \to \infty$, the effect of the Weiner process is negligible in the Langevin equation.
In other words, the fluctuation around the saddle point $m_x = \left\langle \sigma_i^x \right\rangle$ can be ignored.

We can then generate two strategies to employ our approach.
The first strategy is to directly simulate the non-stoquastic Hamiltonian by using the adaptive transverse field:
\begin{enumerate}
\item Perform the quantum Monte Carlo simulation following Eq. (\ref{QMC}) and estimate the value of the expectation of the transverse magnetization. 

To estimate the expectation, we generally require long-time equilibration.
In the case of actual application, we compute the approximate value of the expectation by the empirical average over the whole simulation after the first relaxation.

\item Change the transverse field $\tilde{m}_x$ following the saddle-point solution as
\begin{equation}
\tilde{m}_x = f'(m_x).
\end{equation}
\item Repeat until the physical quantities converge.
\end{enumerate}

\begin{center}
\begin{figure}[t]
\begin{center}
\includegraphics[width=0.4\columnwidth]{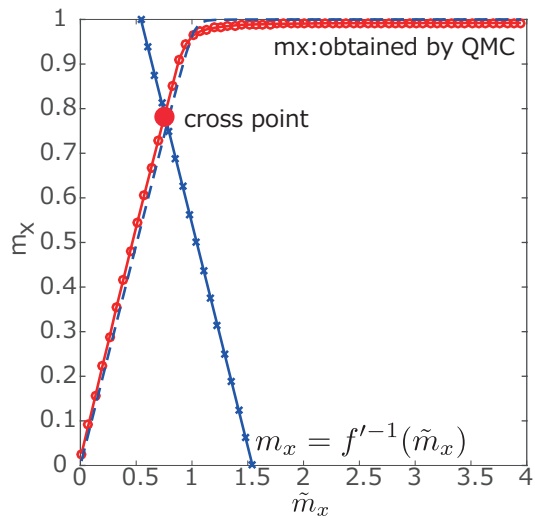}
\end{center}
\caption{Data analysis for the non-stoquastic Hamiltonian.
The red dots are the results of the Ising model, only with the transverse field and the blue dots represent the inverse function as $\tilde{m}_x = f^{-1}(m_x)$.}
\label{demon}
\end{figure}
\end{center}

The second strategy is performed by the data analysis for the results of the ``standard" quantum Monte Carlo simulation of the Ising model, only with the transverse field.
We generate the results obtained by the wide range of the values of the transverse field $\tilde{m}_x$.
Then, we plot the transverse magnetization vs. the transverse field, and we read out its cross point to the curve determined by 
\begin{equation}
m_x = f'^{-1}(\tilde{m}_x),
\end{equation}
where $f'^{-1}(\tilde{m})$ is the inverse function of $f(m)$.
Then, the cross point identifies the realization of the transverse magnetization of the non-stoquastic Hamiltonian.
In the case of the transverse field and the anti-ferromagnetic XX interaction, the inverse function is simply given by the linear function $f'^{-1}(m_x) = (\Gamma - m_x)/\gamma$.
In this example, for the given $\Gamma$ and $\gamma$, we find the cross point and re-plot the realized transverse magnetization and other corresponding thermodynamic quantities vs. the values of $\Gamma$ and $\gamma$, as shown in Fig. \ref{demon}.

\bibliography{main_ver2}
\section*{Acknowledgements (not compulsory)}

The authors would like to thank Hidetoshi Nishimori, and Shu Tanaka for their fruitful discussions.
The present work is funded by the MEXT KAKENHI Grants No. 15H03699, 16H04382, and 16K13849.

\section*{Author contributions statement}

M.O. developed the concept, carried out all the numerical experiments, and prepared the manuscript.

\section*{Additional information}
The author declare no competing financial interests

\end{document}